# Vibration Of A Cantilever Beam In Ambient Fluid

C. Metzger, J. Rodriguez, F. Zypman

Yeshiva University, Department of Physics, New York, NY, US

## ABSTRACT

Here we obtain analytical expressions for the frequency response of a cantilever beam in the presence of ambient fluid. The advantage of our approach, besides its simplicity of use, is that it explicitly contains the viscosity and the density of the ambient fluid. Thus, if measuring the frequency spectrum, the expression can be used in the design of viscometers. Conversely, if the ambient fluid is known, the expression can be used in the design of force gauges such as in Atomic Force Microscopy.

## INTRODUCTION

Our laboratory is interested in modeling the motion of Atomic Force Microscope (AFM) cantilever beams in fluid. We routinely measure interaction forces between the AFM sensor and biological samples that must reside embedded in water. While our interest is not directly in the forces due to the fluid, we need to know them in the process of force reconstruction, to eliminate them and not entangle them with say, electrostatic forces coming directly from the sample charges.

The problem of characterizing the fluid forces is one of a fluid interacting with an elastic moving object, a formidable task impossible to be solved today exactly. In AFM, the problem is compounded by the difficulty of observing the cantilever motion directly since the amplitudes involved are on the order of nanometers. Some laboratories have partially addressed this AFM drawback, but it is today far from routine [1].

Here we tackle the problem by considering a large-scale version of the cantilever. Indeed, we use 30cm-long stainless steel bars that serve as proxies for the AFM cantilever. Since both systems are macroscopic, they satisfy the same dynamics. We first describe the experimental setup and show results of the frequency response in air and liquid. We also develop a simple theory based on Euler-Bernoulli equation to at least qualitatively make sense of the experimental results and trends.

**EXPERIMENTAL**

We use a mechanical oscillator [2] to drive a steel cantilever of mass 9.06 g and dimensions $30 cm \times 12 \mu m \times 1.29 mm$. We place the cantilever as depicted in Figure 1 to be driven on bending, and use a DC power source[3] to run the mechanical oscillator at frequencies in the range 0.1Hz to 4Hz.

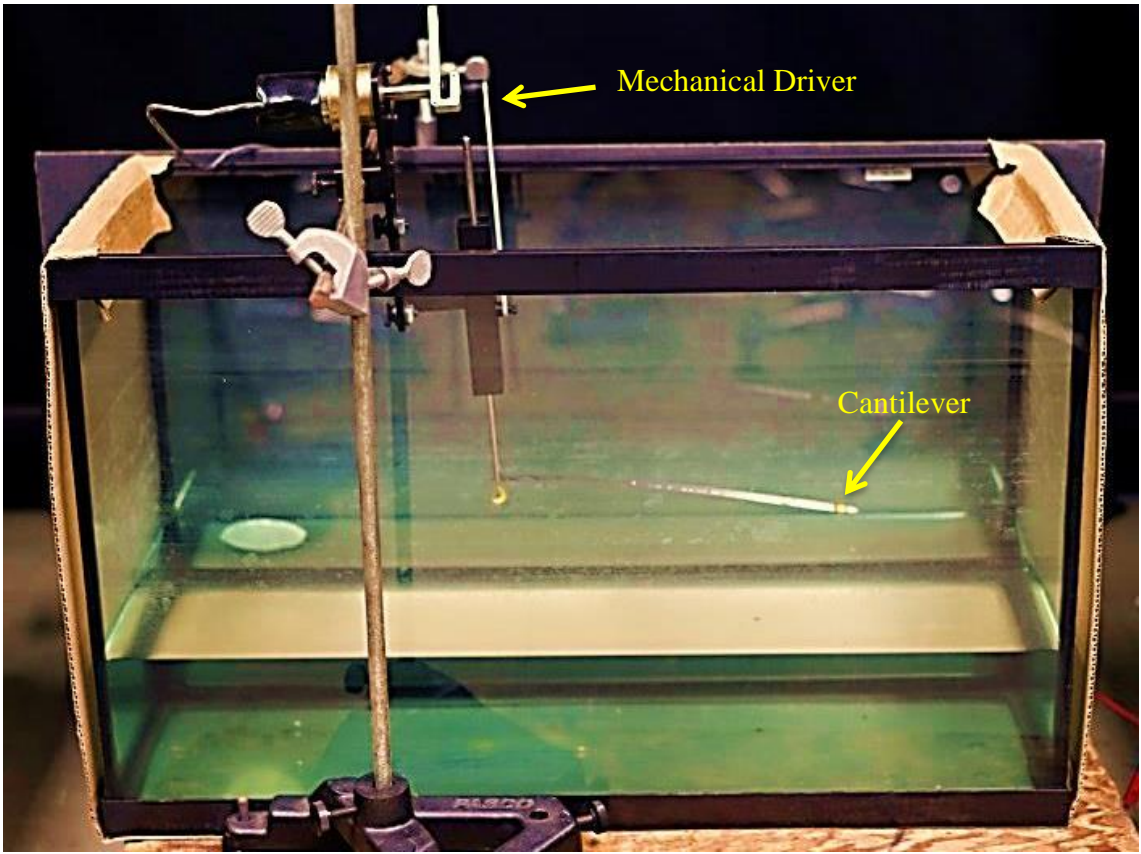

**Figure 1** Experiment showing the cantilever inside the aquarium. It appears as a bright light-reflecting long object in the center of the figure. The mechanical driver consists of the solenoid (black-gold cylinder partially blocked by the stand on the top center-left) and the mechanical lever (vertical black piece with metal rod pointing down and attached to one end of the cantilever).

We filmed two movies of the cantilever in motion both in air and submerged in water on a high resolution camera.[4] From each movie, we extracted the position versus time at both ends: the driven end (left in Figure 1) and the free end (right in Figure 1). The driven end serves as the input amplitude and the free end (the output) varies substantially with frequency. For the water case, we place the cantilever in an aquarium filled with water. The oscillator is above the tank and held in place by a mechanical constraint to prevent the motor's vibrations from skewing the data. The constraint includes both ballast and counter ballast, and a firmly rooted heavy base to minimize outside forces.

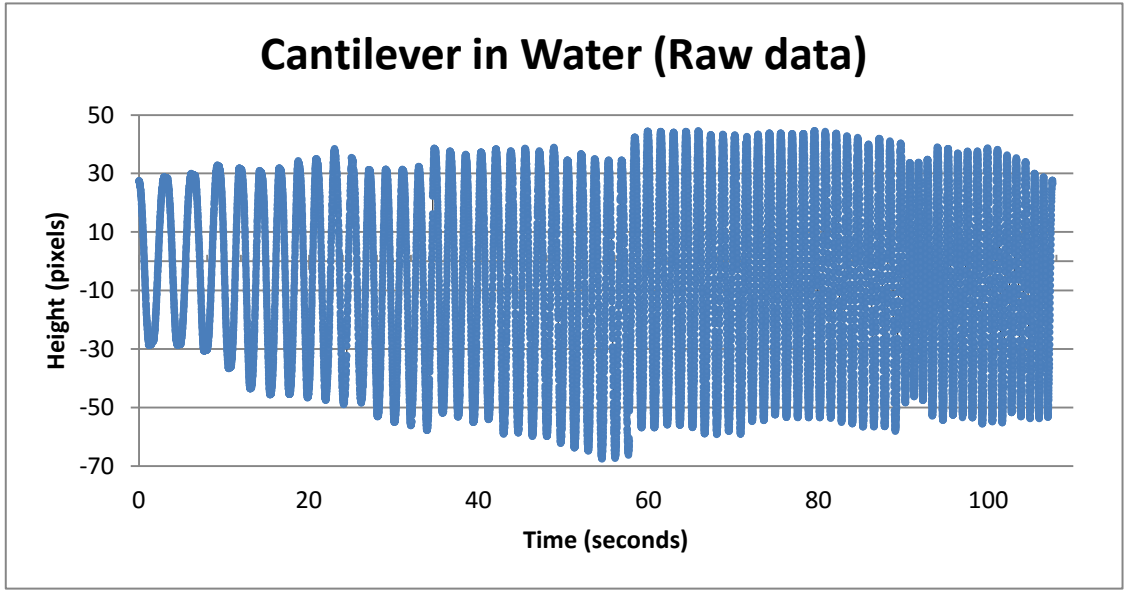

**Figure 2 Raw data in cantilever in water. Each blue dot is a data point. Height of 0 is the height of rod in middle of first oscillation.**

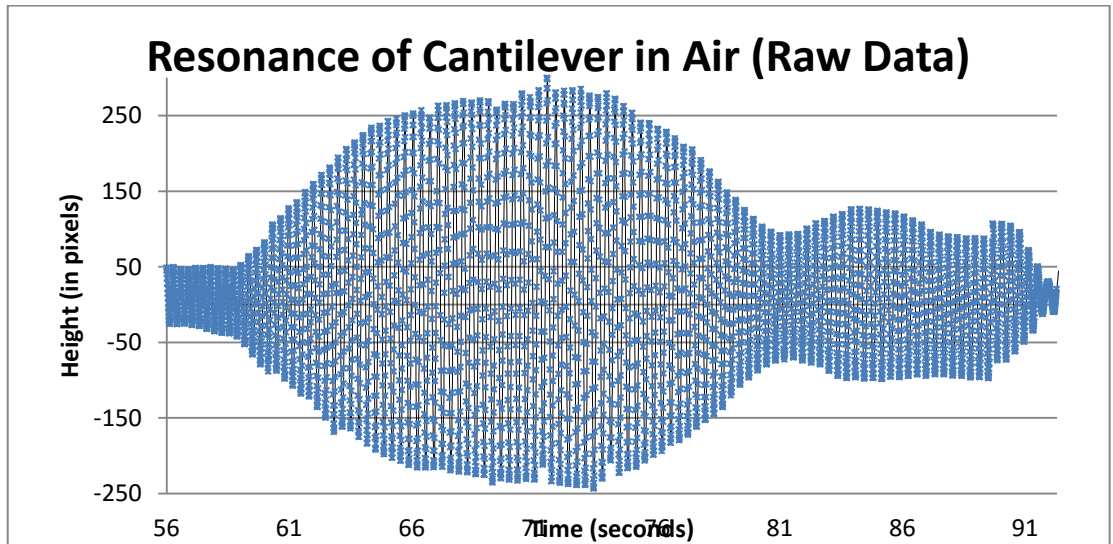

**Figure 3 Raw data for heights of cantilever in air. Each blue dot being a data point. Points connected by the black line to approximate the motion**

.

The recorded video is analyzed with Tracker software[5] (which converts spots in an image to coordinates) to monitor the movement of the edges of cantilever both in air and water. To aid Tracker properly following the right pixels in the frames, we enhanced the contrast by color painting the point of interest. Tracker allows for the data to be saved to disc.

We analyzed the data in Mathematica and Excel and discovered certain trends on the motion of the cantilever as follows.

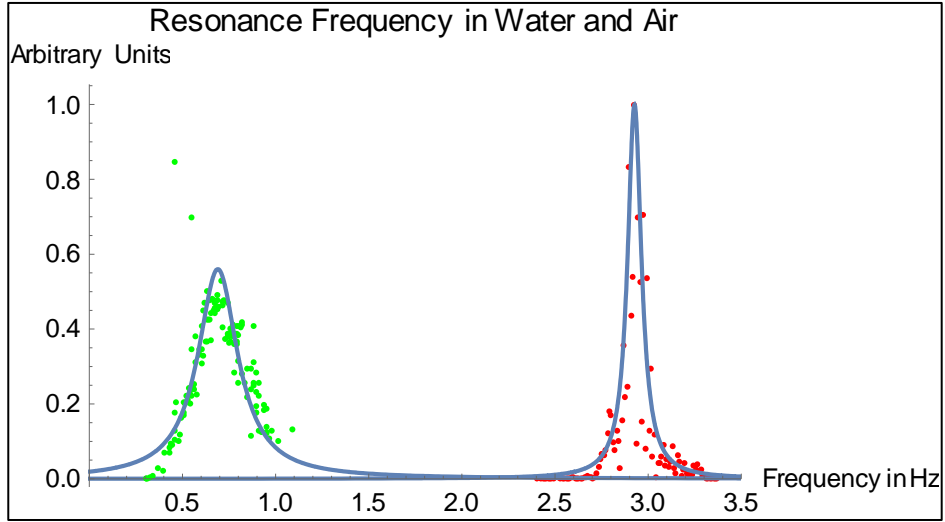

**Figure 4** Experimental versus theoretical resonance frequency of cantilever in water and air

Figure 4 shows maximum amplitude versus frequency of the cantilever in water (green) and in air (red). The blue lines represent the theoretical curves which are determined by the equations for WaterOver (Water Overlaid) and AirOver (Air Overlaid). This equation is given by

$$Theory_{water} = 0.55 * \left( \frac{1}{4} * \frac{(R_{Water})^2}{(f - f_{0Water})^2 + \left(\frac{1}{2} R_{Water}\right)^2} \right) \quad (1)$$

$$Theory_{air} = \left( \frac{1}{4} * \frac{(R_{Air})^2}{(f - f_{0Air})^2 + \left(\frac{1}{2} R_{Air}\right)^2} \right) \quad (2)$$

The $R_{water}$ is the density of the water, $R_{air}$ that of air. With $f_{0Air}$ and $f_{0Water}$ being the frequency where the cantilever reached maximum resonance in air and water respectively. Experimentally we found that $R_{Water}=0.31$; $R_{Air} = 0.09$; $X_{0Air}=2.926732441$ Hz; $X_{0Water}=.699193$ Hz;

**THEORETICAL**

We begin from the Euler-Bernoulli equation which gives a good representation of the dynamics of the system, particularly for the lowest modes of vibration -our experiments concentrate on measurements of the first mode since higher modes in water were not observed, presumably due to high dissipation loss.

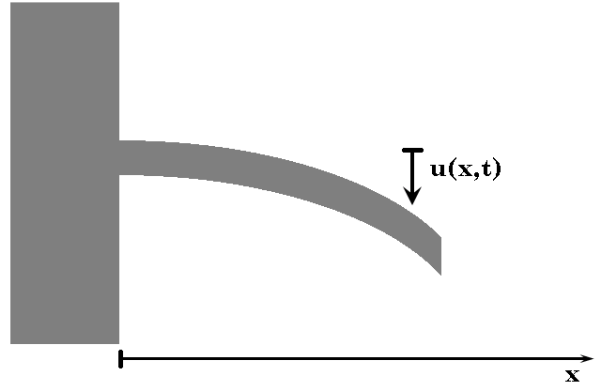

**Figure 3** Diagram representation of the cantilever.

The cantilever's dimensions are dominated by its length L -measure along the x-axes. The width a, thickness w, and the displacement $u(x,t)$ are all much smaller than L. In our setup $u(x,t)$ is measured with respect to the horizontal line.

Euler-Bernoulli equation describes the dynamics of the cantilever,

$$\frac{\partial^4 u(x,t)}{\partial x^4} + \frac{\partial^2 u(x,t)}{\partial t^2} = 0 \qquad (1)$$

where the first term describes the elasticity of the cantilever and the second is the inertia. Equation (1) is written in dimensionless form, where both $u(x,t)$ and $x$ are measured in units of the length L, and the time is measured in units of $t_0 = \sqrt{\frac{\rho L^4}{\Re^2 E}}$, where $\rho$ is the density of the cantilever, $E$ its Young's modulus, and $\Re$ the radius of gyration of the cross-sectional area with respect to and axis perpendicular to the plane of the page through the center of the cross section.

In real situation, there also exists energy dissipation. This may come form internal motion, or from the relative sliding of the cantilever with respect to the ambient fluid (air, water, for example). The fluid mechanics problem of a fluid moving relative to a moving solid is difficult, and many attempts have been made in the AFM community to lump the effects into a simple dissipation term. This program is ongoing and well beyond the intent of this study. We incorporate an ad hoc dissipation term proportional to the local velocity of the cantilever whose strength can be tuned to model fluid environments: the higher the strength, the higher the dissipation, which can be qualitatively thought of as a measure of viscosity.

Therefore we consider a dissipative version of the Euler-Bernoulli equation thus,

$$\frac{\partial^4 u(x,t)}{\partial x^4} + \frac{\partial^2 u(x,t)}{\partial t^2} + \Gamma(u(x,t)) = 0 \qquad (2).$$

where $\Gamma(u(x,t))$ is a functional of $u(x,t)$, whose exact form has been studied analytically for spheres [6]. For cylindrical shapes, exact solutions are known only for rigid vibrations [7] of circular cylinders. Here we use $\Gamma(u(x,t))$ as developed previously to study high viscous liquids [8]. The functional itself is cumbersome, but its Fourier Transform is simple and we will introduce it explicitly below.

As we are interested in monochromatic excitations, we consider the following boundary conditions for equation (2):

$$[u(x,t)]_{x=0} = e^{i\omega t} \qquad (3a)$$

$$\left[\frac{\partial u(x,t)}{\partial x}\right]_{x=0} = 0 \qquad (3b)$$

$$\left[\frac{\partial^2 u(x,t)}{\partial x^2}\right]_{x=1} = 0 \qquad (3c)$$

$$\left[\frac{\partial^3 u(x,t)}{\partial x^3}\right]_{x=1} = 0 \qquad (3d)$$

Equation (3a) establishes a vibration of angular frequency $\omega$ and unit amplitude at held extreme. The choice of unit amplitude is possible since the system is linear -any other choice of amplitude would appear as a simply multiplicative factor along the cantilever. Equation (3b) fixes the slope of the cantilever horizontal at the held extreme of the cantilever.

No torques act on the free end of the cantilever; this is given by equation (3c).

Finally, equation (3d) represents that no forces act at the free and of the cantilever.

Given condition (3a), and the linearity of equation (2), we seek solutions of the form

$$u(x,t) = f(x)e^{i\omega t} \quad (4).$$

With equation (4), equations (2) and (3) now become,

$$\frac{d^4 f(x)}{dx^4} - (\omega^2 + g(\omega)) f(x) = 0 \quad (5)$$

$$f(0) = 1 \quad (6a)$$
$$f'(0) = 0 \quad (6b)$$
$$f''(1) = 0 \quad (6c)$$
$$f'''(1) = 0 \quad (6d)$$

The new function $g(\omega)$ is (times $f(x)$) the time Fourier Transform of $\Gamma(u(x,t))$ and takes the form

$$g(\omega) = \frac{9\omega^2}{2}\sqrt{\frac{2\eta\rho_F}{\omega}}\left(1 + \frac{2}{9}\sqrt{\frac{\rho_F \omega}{8\eta}}\right) - 18i\eta\omega\left(1 + \sqrt{\frac{\rho_F \omega}{8\eta}}\right) \quad (7)$$

where $\eta = \frac{(\text{dynamic\_viscosity})t_0}{\rho a^2}$ and $\rho_F = \frac{\text{fluid\_density}}{\rho}$ [8].

Equation (5) has solutions in terms of trigonometric and hyperbolic trigonometric functions of argument $\left(\omega^2 + g(\omega)\right)^{1/4} x$. For low viscosities, resonances appear as large amplitudes $f(x)$ close to the frequencies of normal modes.

Our experiments, however, measured only the motion of the cantilever close to the first resonance, consequently we explore a simpler representation of the solution to (5) for the lowest frequency. This corresponds to a one-degree-of-freedom system with a nodeless $f(x)$. Thus, we propose a solution

$$f(x) = \sum_{n=0}^{4} C_n x^n \qquad (8)$$

First, we impose the four boundary conditions (6), which reduce the five coefficients in (8) to (for example) just $C_2$. The solution (8) that satisfies (6) then is

$$f(x) = 1 + C_2 \left( x^2 - \frac{2}{3} x^3 + \frac{1}{6} x^4 \right) \qquad (9)$$

Due to the boundary conditions (6), the solution $f(x)$ goes from a horizontal close to $x = 0$ to a straight line close to $x = 1$ – most of the kinetic energy is in the region close to $x = 1$. To obtain an approximation to the solution, we substitute the right-hand-side of equation (9) into equation (5) and find $C_2$ by imposing that (5) is exactly correct at $x = 2/3$, closer to the right end of the cantilever.

This procedure gives explicitly,

$$C_2 = \frac{243}{4} \frac{\omega^2 + g(\omega)}{243 - 17\omega^2 - 17g(\omega)} \qquad (10)$$

Then, the quantity monitored experimentally oscillates harmonically with amplitude

$$f(1) = \left| \frac{7776 + 1643\omega^2 + 96(17\omega^2 - 243) - 4(695\omega^2 + 486) + 495g(\omega)}{48(17g(\omega) + 17\omega^2 - 243)} \right| \qquad (11)$$

where $|\ |$ stands for the complex modulus.

To gain qualitative insight into the solution in the equation (11), we apply equation (11) (together with equation (7)) to air and water.

Figure 4 shows the results for such calculations, for which we used $997 Kg/m^3$ for the density of water, $0.8 Kg/m^3$ for the density of air, $8000 Kg/m^3$ for the density of steel, $1.86\times10^{-5}$ Pa.s for the dynamic viscosity of air, $8.9\times10^{-4}$ Pa.s for the dynamic viscosity of water, and a cantilever width of 1.5 cm. In addition, as described in [8] we used a (single) form factor of 30 to account for water drag and a consequently larger effective cantilever. The results agree very well with the experiments. We see in Figure 4 that the originally sharp peak in water becomes broad and slower in the presence of water.

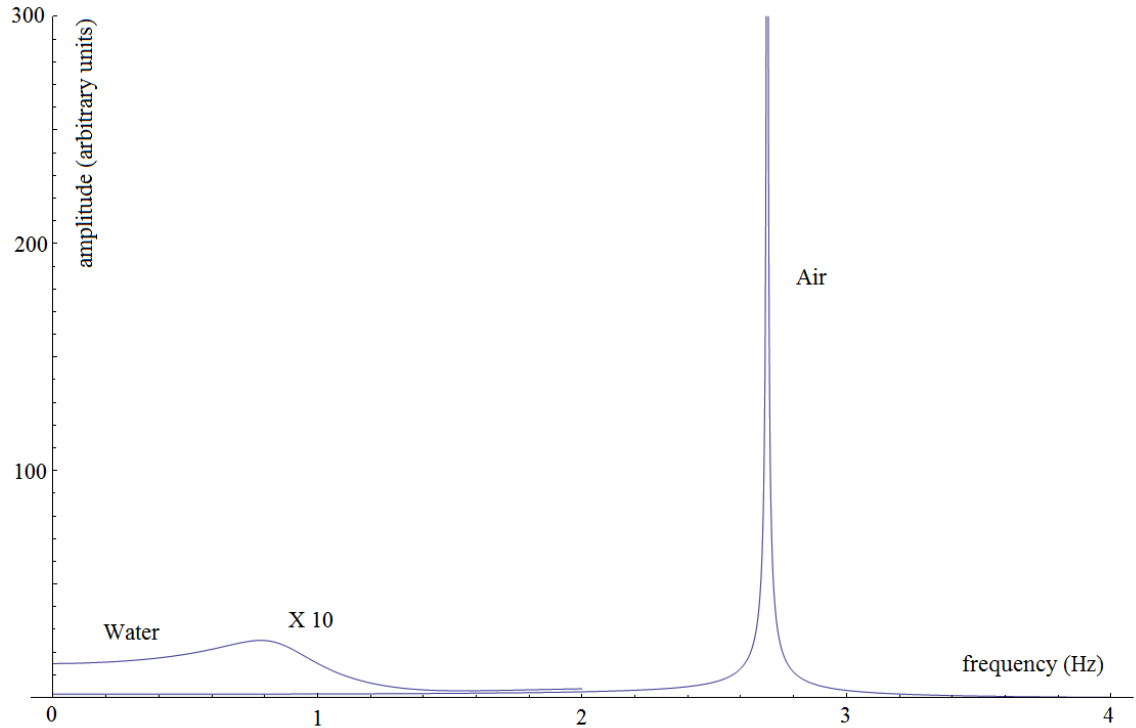

**Figure 4**. Theoretical frequency response curve results as calculated using equation (11) for water and air. The frequency response amplitude for water would be too small to easily see it in this scale so we magnified it by a factor of 10.

**CONCLUSIONS**

We have done scaled up measurements of AFM cantilevers to gain insight into the effect of fluids on the dynamics of the cantilever. More importantly, we tested whether a theory based on linked chains can correctly describe the experimental results. The result is positive and should be of practical use when designing AFM computer algorithms that require knowledge of the motion of the cantilever in fluids.

ACKNOWLEDGMENT

Project funded by the National Science Foundation grant #CHE-1508085